\documentclass{revtex4}
\usepackage{psfig}
\usepackage{subfigure}
\usepackage{graphicx}
\usepackage{doublespace}
\usepackage{amsmath}
\begin{document}

\doublespace

\renewcommand{\baselinestretch}{2}
%\draft

%\onecolumn

\doublespace

\author{Francesco Sorrentino${}^{*\ddagger}$, Edward Ott${}^{*}$ }
\affiliation{${}^*$ Institute for Research in Electronics and Applied Physics, Department of Physics, and Department of Electrical and Computer Engineering, University of Maryland, College Park, Maryland 20742\\${}^\ddagger$ Universit{\`a} degli Studi di Napoli Parthenope, 80143 Napoli, Italy}

\begin{abstract}
We  address the issue of how to identify the equations of a largely unknown chaotic system from knowledge of its state evolution. The technique can be applied to the estimation of parameters that drift slowly with time. To accomplish this, we  propose an adaptive strategy that aims at synchronizing the unknown real system with another system whose parameters are adaptively evolved to converge on those of the real one. Our proposed strategy is tested to identify the equations of the Lorenz, and the R\"ossler systems. We also consider the effects of measurement noise and of deviation of our fitting model from consistency with the true dynamics.

\end{abstract}

\title{Using synchronization of chaos to identify the dynamics of unknown systems}
%\pacs{05.45.Xt,87.18.Sn,89.75.-k}
\maketitle

\textbf{In this paper, we present a new technique to estimate the equations of an unknown chaotic system. We assume that suitable but rather minimal information is available about the form of the system equations and we use the observed dynamics to estimate an approximation to the unknown system. We envision that our findings could be useful in situations in which the equations governing the dynamics of a given unknown chaotic system are to be identified or in situations where system parameters drift with time.
}

\section{Introduction}

Systems of nonlinear differential equations are often used to study the dynamics of real world systems. For instance, the Hindmarsh-Rose equation \cite{HR} is widely considered to be a reasonable model of the firing/bursting behavior in real neurons, and has been shown to be chaotic for a certain range of its parameters. Another example is that of the Chua system \cite{Matsu,Matsu2} that models the dynamics of a simple electronic circuit for which the emergence of chaos has been observed both in simulations and in experiments.
A general question that often arises is whether such a system of differential equations replicates the dynamics of a given real system that it is meant to model. % quantitatively rather than just qualitatively.

In a recent paper \cite{CRMR}, it has been shown that it is possible to synchronize a real experimental system with a system of differential equations simulated on a computer by coupling the two. This is only possible if the system simulated on the computer does a sufficiently good job of quantitatively replicating the dynamics of the experimental system \cite{Crutch}. An interesting implication of this is that it is possible to quantitatively evaluate the degree to which a given mathematical model replicates the dynamics of a real system via its capability to synchronize with the real system when the two are coupled.
Here we adopt a different but related application of synchronism of coupled chaotic systems. In particular, we consider that the real system is largely unknown, and we propose an adaptive strategy that, by using synchronization, is able to obtain a set of (nonlinear) differential equations that describes it. Furthermore, in the case that the fitting function basis for our model is not consistent with the true system dynamics, we will try to answer the question of how well the obtained model is able to forecast the true system future behavior.

Refs. \cite{Nij,Nij2} have outlined the connection between the problem of synchronization of dynamical systems and the problem of the design of an observer to reconstruct the state of an unknown system. The idea of using synchronization or control for parameter and model identification has previously been presented in \cite{So:Ott:Day} and subsequently in \cite{Abarbanel,Abarbanel2,Yu:Parlitz}. Some recent papers have appeared on reconstructing the state and parameters of the Chua system \cite{Chua_id1,Chua_id2}. Here we will address the issue of how to identify the equations of a largely unknown generic system from knowledge of its state evolution. In Sec. II, the problem is introduced and an adaptive strategy, based on the minimization of appropriately defined potentials (see e.g., \cite{SOTT}), is presented. In Sec. III, our proposed strategy is tested to identify the equations of the R\"ossler, and the Lorenz systems. In Sec. IV, we will take into account the effects of measurement noise. In Sec. V, we will address the case that the fitting function basis for our model is not consistent with the true system equations and for this case, we will study the capability of the obtained approximate model to forecast the evolution of the true system.

%As a first example we consider that our unknown experimental system is a R\"ossler system $\dot{\bf{x}}=F({\bf{x}})$, where ${\bf{x}}$ is three dimensional vector, ${\bf{x}}=(x_1,x_2,x_3)$ and,

\section{Formulation}

We consider the general example, $\dot{\bf{x}}=F({\bf{x}})$, with ${\bf{x}}=(x_1,x_2,...,x_n)^T$ and $F({\bf{x}})=[f_1({\bf{x}}),f_2({\bf{x}}),...,f_n({\bf{x}})]^T$, where
\begin{equation}
f_i({\bf{x}})=\sum_{j=1}^{n} \sum_{k=j}^{n} a_{ijk} x_j x_k+ \sum_{j=1}^{n} b_{ij} x_j +c_i.  \label{TO}
\end{equation}
That is, $f_i(\bf{x})$ is a degree 2 polynomial. Note that the specification of $F({\bf x})$ from (\ref{TO}) involves $M=[(n^2+n)/2+n+1]n$ parameters. For example, in the case of the R\"ossler system (to be used later in our numerical experiments), $n=3$, and
\begin{equation}
F({\bf{x}})=\left[\begin{array}{c}
    -x_{2}-x_{3} \\
    x_{1} + 0.165 x_{2} \\
    0.2+ (x_{1}-10 )  x_{3}
  \end{array} \right], \label{F}
\end{equation}
and all the coefficients ($a_{ijk},b_{ij}$, and $c_i)$ are zero, except for $a_{313}=1$, $b_{12}=-1$, $b_{13}=-1$, $b_{21}=1$, $b_{22}=0.165$, $b_{33}=-10$, $c_3=0.2$. The Lorenz system could be recast in similar terms. % of the system (\ref{G}), as well.
The Hindmarsh-Rose system, has an analogous structure, but with higher order powers of the state variables as well (one term is third order).

In this paper, we will be mainly concerned with the case that the chosen form for our model system is a degree 2 polynomial, that is as in Eq. \ref{TO}. Nonetheless the extension to the case of higher order polynomials or to a different fitting functions basis is straightforward.

%Note that in (\ref{G}), we have chosen each one of the functions $f_1,f_2,f_3$ to be a multivariate polynomial of the state variables $x_1,x_2,x_3$ of \emph{degree} $N=2$ (by this here we mean that for each term in (\ref{G}) of the type $x_1^{\alpha} x_2^{\beta} x_3^{\gamma}$, $\alpha+\beta+\gamma$ must be less or equal $2$).

%The system of equation in (\ref{G}), can be rewritten in a more compact way as follows,
%\begin{equation}
%\dot{x}_i=f_i(x_1,x_2,...,x_n)=\sum_{j=1}^{n} \sum_{k=j}^{n} a_{ijk} x_j x_k+ \sum_{j=1}^{n} b_{ij} x_j +c_i, \qquad i=1,...,n \label{G1}
%\end{equation}
%

We assume that the exact system (2) is unknown, but it is known that the system is of the form $\dot{\bf{x}}=F({\bf{x}})$, with $\bf{x}$ $n$-dimensional, and $F({\bf x})$ expressible, or approximately expressible, in terms of degree 2 polynomials in $\bf{x}$ as discussed above. It is then appropriate to try to model the dynamics of the true system by  $\dot{\bf{x'}}=F'({\bf{x'}})$, with
\begin{equation}
\dot{x'}_i=f'_i(x'_1,x'_2,...,x'_n)=\sum_{j=1}^{n} \sum_{k=j}^{n} a'_{ijk} x'_j x'_k+ \sum_{j=1}^{n} b'_{ij} x'_j +c'_i. \qquad i=1,...,n. \label{G2}
\end{equation}
Our goal is then to obtain good estimates $a'_{ijk}$, $b'_{ij}$, and $c'_i$ of the true coefficients $a_{ijk}$, $b_{ij}$, and $c_i$. We assume that, although $F({\bf x})$ is unknown, we do have access to good measurements of the evolving `experimental' system. To accomplish our goal, we envision coupling the true system to the model system (3). Our approach will be to formulate an adaptive procedure which adjusts the model coefficients $a'_{ijk}$, $b'_{ij}$, and $c'_i$ in such a way as to achieve synchrony. It is then hoped that, when this has been accomplished, the model coefficients will be a good approximation to the corresponding coefficients of the real system.

We perform a one way diffusive coupling from the true system to the model, as follows
\begin{equation}
\dot{\bf{x'}}=F'({\bf{x'}})+\Gamma(H({\bf{x}})-H({\bf{x'}})). \label{dyn}
\end{equation}
%\begin{equation}
%\dot{\bf{x'}}=F'({\bf{x'}})+\gamma(F({\bf{x}})-F'({\bf{x'}})).
%\end{equation}
%Note that for $\gamma=1$, this is equivalent to
%\begin{equation}
%\dot{\bf{x'}}=F({\bf{x}}).
%\end{equation}
The quantity $H$ is in general an $m\leq n$ vector of $m$ observable scalar quantities that are assumed to be known functions of the system state $ {\bf{x}}(t)$. $\Gamma$ is an $m \times n$ constant coupling matrix. For our numerical experiments we will assume that: (i) $H({\bf{x}})={\bf{x}}$; (ii) $\Gamma= \gamma I_n$, where $\gamma$ is a scalar quantity, and $I_n$ is the identity matrix of dimension $n$. Note that our strategy requires that, when the system parameters are correctly identified, i.e., $a'_{ijk}=a_{ijk}$, $b'_{ij}=b_{ij}$, $c'_i=c_i$,  $\gamma$ belongs to the range for which the synchronized solution ${\bf{x}}(t)={\bf{x}}'(t)$ is stable with respect to infinitesimal perturbations \cite{Pe:Ca}.

We now introduce the following potentials,
\begin{equation}
\Psi_i=<[\dot{x}_i-f'_i(x'_1,x'_2,...,x'_n)]^2>_{\nu}, \qquad i=1,2,...,n,   \label{P}
\end{equation}
where $<G(t)>_{\nu}$ denotes the sliding exponential average $\nu \int^t e^{-\nu (t-t')} G(t') dt'$. Note that $\Psi_i$ is a function of time and also of the coefficients
$a'_{ijk}$, $b'_{ij}$, and $c'_i$. Also note that if $x_1(t),x_2(t),...,x_n(t)$ are chaotic, then $f_1(x_1(t),x_2(t),...,x_n(t)),f_2(x_1(t),x_2(t),...,x_n(t)),...,f_n(x_1(t),x_2(t),...,x_n(t))$ vary chaotically, as well. Furthermore, we point out that the exponential averaging operation $<G(t)>_{\nu}$ is the same as low-pass filtering of $G(t)$, using a first order filter of bandwidth $\nu$. The potential (5) satisfies $\Psi_i \geq 0$ and has a minimum value of zero. A sufficient condition for the potentials in (\ref{P}) to be zero is
\begin{equation}
{x}_i(t)={x'}_i(t), \qquad i=1,2,...,n,  \label{COND1}
\end{equation}
and
\begin{equation}
\begin{split}
a'_{ijk}= a_{ijk}, \qquad & i,j=1,...,n;  k=j,...n,\\
b'_{ij}= b_{ij}, \qquad & i,j=1,...,n, \\
c'_i=c_i, \qquad & i=1,..,n. \label{COND2}
\end{split}
\end{equation}
% these conditions are equivalent to,
%\begin{subequations}
%\begin{align}
%{x}_1(t)={x'}_1(t) \\
%{x}_2(t)={x'}_2(t) \\
%{x}_3(t)={x'}_3(t),  \label{CONDe1}
%\end{align}
%\end{subequations}
Equation (\ref{COND1}) corresponds to synchronization. Equation (\ref{COND2}) corresponds to a correct identification of the (unknown) system parameters. Due to the chaotic nature of the true system, it should typically be the case that this sufficient condition for the potential to be zero is also necessary.

We note that in the case in which our proposed model form is consistent with the dynamics of the true system (e.g., they are both expressed by a degree two-polynomial as assumed in Eqs. ($\ref{TO}$) and ($\ref{G2}$)), there is only one possible choice of $a'_{ijk},b'_{ij},$ and $c'_i$ that minimizes the potentials (\ref{P}), namely , $a'_{ijk}=a_{ijk}$, $b'_{ij}=b_{ij}$, $c'_i=c_i$. This follows from the principle that two polynomials are equal only if all their coefficients are equal and from the fact that (\ref{P}) implies a one to one correspondence between the model state variables and those of the true system.

We propose to adaptively evolve the estimates of the parameters $a'_{ijk},b'_{ij},c'_i$ in time, according to the following gradient descent relations:
\begin{subequations}
\begin{align}
\frac{d {a'}_{ijk}(t)}{dt}=& -\beta_a \frac{\partial \Psi_i}{\partial {a'}_{ijk}}, \\
\frac{d {b'}_{ij}(t)}{dt}=& -\beta_b \frac{\partial \Psi_i}{\partial {b'}_{ij}}, \\
\frac{d {c'}_{i}(t)}{dt}=& -\beta_c \frac{\partial \Psi_i}{\partial {c'}_{i}},  \label{CONDe1}
\end{align}
\end{subequations}
$\beta_a,\beta_b,\beta_c>0$. Our hope is that $a'_{ijk}(t)$, $b'_{ij}(t)$, and $c'_i(t)$ will converge under this evolution to the true parameter values, $a_{ijk},b_{ij},c_i$.

First we consider (8a). Let $({f'_i})_{jk}$ denote $f'_i(x'_1,x'_2,...,x'_n)$ evaluated at $a'_{ijk}=0$, %and ${\tilde{f'}_1}^{a_{12}}=f'_1-{f'_1}^{a_{12}}$.
\begin{equation}
f'_i(x'_1,x'_2,...,x'_n)={a'_{ijk}} x'_j x'_k+ ({f'_i})_{jk},
\end{equation}
%\begin{equation}
%{f'_i}^2(x'_1,x'_2,...,x'_n)={a'_{ijk}}^2 {x'_j}^2 {x'_k}^2+  2 ({f'_i})_{jk} a'_{ijk} x'_j x'_k + {({f'_i})_{jk}}^2 .
%\end{equation}
Substituting this into the right hand side of Eq. (8a), we obtain,
\begin{equation}
\frac{d {a'}_{ijk}(t)}{dt}= - 2 \beta_a <a'_{ijk} {x'_j}^2 {x'_k}^2+ {({f'_i})_{jk}}x'_j x'_k-\dot{x}_i x'_j x'_k>_{\nu}.
\end{equation}
%If $a'_{12}$ evolves on a time scale that is large compared to $\nu^{-1}$, we have,
%\begin{equation}
%\frac{\partial \Psi_1(t)}{\partial {a'}_{12}(t)}\simeq-2 \beta_1(a'_{12} <(x'_1 x'_2)^2>_{\nu}+ <({f'_1}^{a_{12}}-f_1)x'_1 x'_2>_{\nu}).
%\end{equation}
Similarly letting $({f'_i})_j$ denote $f'_i(x'_1,x'_2...,,x'_n)$ evaluated at $b'_{ij}=0$, Eq. (8b) gives
\begin{equation}
\frac{d {b'}_{ij}(t)}{dt}= -2 \beta_b <b'_{ij} {x'_j}^2+ {({f'_i})_j} x'_j-\dot{x}_i x'_j >_{\nu}.
\end{equation}
Finally, we consider relation (8c) with $(f'_i)$ denoting $f'_i(x'_1,x'_2,...,x'_n)$ evaluated at $c'_i=0$. Then
\begin{equation}
\frac{d {c'}_{i}(t)}{dt}= -2 \beta_c <c'_{i} + (f'_i)-\dot{x}_i >_{\nu}.
\end{equation}

In this paper we consider the case where $\beta_{a,b,c}$ are very large. For this situation the solutions $a'_{ijk}$, $b'_{ij}$, $c'_i$ rapidly converge to the minimum of the potentials (which is zero). Thus we can set the the averages $<...>_{\nu}$  on the right hand side of Eqs. (10)-(12) to zero. We further consider that the average $<...>_{\nu}$ is performed over a time scale $\nu^{-1}$ that is much larger than the time scale $T_s$ for variation of ${\bf x}(t)$, in which case $a'_{ijk}(t)$, $b'_{ij}(t)$, $c'_i(t)$ vary slowly compared to ${\bf x}(t)$. Under these conditions, (8), (10), (11), and (12) then yield
\begin{subequations}
\begin{align}
\sum_{l=1}^n \sum_{m=l}^n a'_{ilm} {<x'_l x'_m x'_j x'_k>_{\nu}}+ \sum_{l=1}^{n} b'_{il} <x'_l x'_j x'_k>_{\nu} +c'_i <x'_j x'_k>_{\nu} =& <\dot{x}_i x'_j x'_k>_{\nu}, \qquad & j=1,...,n; \quad k=j,...,n, \\ %a'_{ijk} {<(x'_j x'_k)^2>_{\nu}} +
 \sum_{l=1}^n \sum_{m=l}^n a'_{ilm} <x'_l x'_m x'_j>_{\nu} + \sum_{l=1}^n b'_{il} <{x'_l}{x'_j}>_{\nu}+ c'_i <x'_j>_{\nu}= & <\dot{x}_i x'_j>_{\nu}, \qquad & j=1,...,n, \\
 \sum_{l=1}^n \sum_{m=l}^{n} a'_{ilm} <x'_l x'_m>_{\nu} + \sum_{l=1}^{n} b'_{il} <x'_l>_{\nu}+  {c'_i}= & <\dot{x}_i>_{\nu}.
\end{align}
\end{subequations}

Equations (13) constitute a system of $M=[(n^2/2)+(3n/2)+1]n$ linear equations for the $M$ quantities $a'_{ijk}$, $b'_{ij}$, and $c'_i$.
The coefficients of the quantities to be solved for, as well as the driving factors on the right hand sides of Eqs. (13), are all of the form of an average $<(...)>_{\nu}$, where for the coefficients of the unknown $(...)$ is a product of $x'$ terms, while for the driving factors $(...)$ involve the time derivative $\dot{\bf x}$ of the observed experimental system state. In practice, it is inconvenient to explicitly calculate the integrals for these quantities, in terms of the form,
\begin{equation}
I(t)=<G(t)>_{\nu}= \nu \int^t e^{-\nu(t-t')} G(t') dt',
\end{equation}
at every time step. Instead we shall use the fact that $I(t)$ satisfies the differential equation,
\begin{equation}
\frac{dI}{dt}+ \nu I= \nu G(t)
\end{equation}
and obtain $I(t)$ as a function of time by solving (15).
Thus our adaptive system for finding estimates of the quantities, $a_{ijk}, b_{ij}, c_i$, is Eq. (4) for ${\bf x'}(t)$, Eqs. (13) for $a'_{ijk}(t)$, $b'_{ij}(t)$, and $c'_i(t)$, where the various terms in (13) that are of the form $I(t)=<G(t)>_{\nu}$ obtained by integrating (15). If our procedure works, the time evolutions of $a'_{ijk}(t)$, $b'_{ij}(t)$, $c'_i(t)$ will converge to  $a_{ijk}$, $b_{ij}$, $c_i$ with increasing $t$.

{In this paper, we focus on a case in which the true system equations are linear in the unknown coefficients (e.g., they are expressible in forms of given degree polynomials as in (1)). Under this assumption, the minimization of the potentials  can be achieved by solving a system of linear equations as in (13), which in practice, is simpler than solving the system of differential equations (10-12).  We note, however, that, our strategy can also be employed, when the true system equations (1) are nonlinear in the unknown coefficients. For such a case, it may be impossible to obtain a simple unique solution for the unknown coefficients as in (13), and integration of the gradient descent differential equations (e.g., Eqs. (10-12)) is a potentially useful approach.
}

\section{NUMERICAL EXPERIMENTS}

\subsection{An experiment with the R\"ossler system}

We now present numerical experiments testing the above strategy for the example in which the unknown system is the R\"ossler system described by Eq. (\ref{F}). The system in  (\ref{F}) is evolved starting with a random initial condition on the attractor $x(0),y(0),z(0)$, while the system in (\ref{G2}) is evolved starting from a perturbed initial condition,
\begin{equation}
x'_{1}(0)= x_1(0) +  \rho_1 \epsilon_{x}, \quad  x'_{2}(0)= x_2(0) +  \rho_2 \epsilon_{y}, \quad  x'_{3}(0)= x_3(0) +  \rho_3 |\epsilon_{z}|, \label{IC}
\end{equation}
where $\epsilon_{x}, \epsilon_{y}$ and $\epsilon_{z}$ are zero-mean independent random numbers of unit variance drawn from a normal distribution; $\rho_1=7.45,\rho_2=7.08,\rho_3=4.25$ are the standard deviations of the time evolutions of the state components of the true system.
We also need to specify the initial conditions for Eq. (15) for the internal variables of the form $<...>_{\nu}$. For our experiment, these are all initially set equal to random numbers drawn from a gaussian distribution with zero mean and standard deviation equal to $0.1$. $\nu$ is equal to $10^{-2}/2$, so that the sliding exponential time average $<...>_{\nu}$ is performed over a time $\nu^{-1}=200$ that is long compared to the characteristic time $T_s$ for the evolution of a R\"ossler system. We estimate the latter time  to be about $T_s\simeq 6$, which is the average measured time interval between two consecutive peaks of $x_1(t)$.
Numerical results are shown in Figs. 1-2.
Figure 1  shows the time evolutions of $x_1(t)$ and $x'_1(t)$ (respectively, $x_2(t)$ and $x'_2(t)$, $x_3(t)$ and $x'_3(t)$) at the beginning and at the end of the simulation (note that at the end of the run, synchronization is attained for all the three state variables). Figure 2 shows the time evolution of \emph{some} of the estimated parameters (in red), when compared to the corresponding \emph{true values} in system (\ref{G2}) (black dotted lines). The values of all the $M$ estimated parameters $a'_{ijk}(t)$, $b'_{ij}(t)$, $c'_i(t)$ at the end of the run ($t=10^4$) are reported in Table 1. It is seen that both the coefficients that have {true values} equal zero and those that have {true values} different from zero are accurately estimated. %to within a certain degree of approximation.

\begin{figure}[h]
\centerline{\psfig{figure=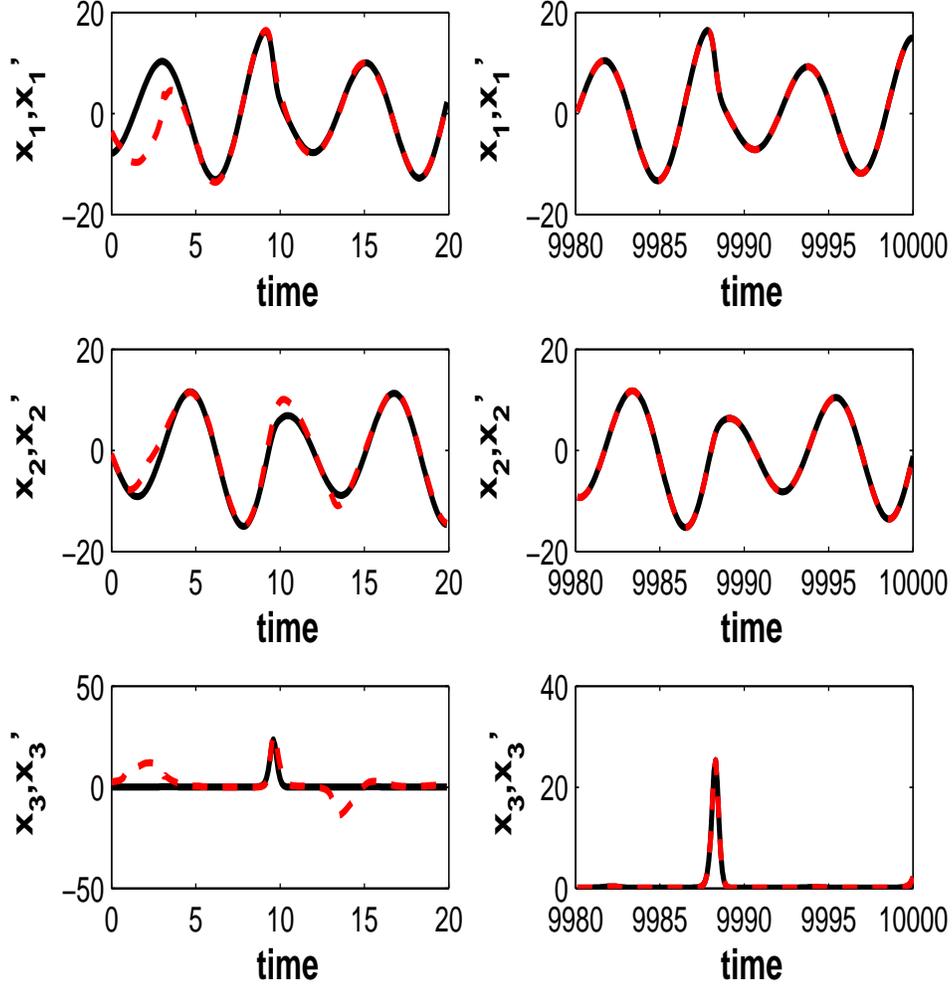,width=14cm,height=15cm}}
\caption{The figure shows the time evolutions of $x_1(t),x_2(t),x_3(t)$ (in black) and of $x'_1(t),x'_2(t),x'_3(t)$ (in red) at the beginning and at the end of the simulation. $\nu=10^{-2}/2$, $\gamma=2$. \label{S}}
\end{figure}
\begin{figure}[h]
\centerline{\psfig{figure=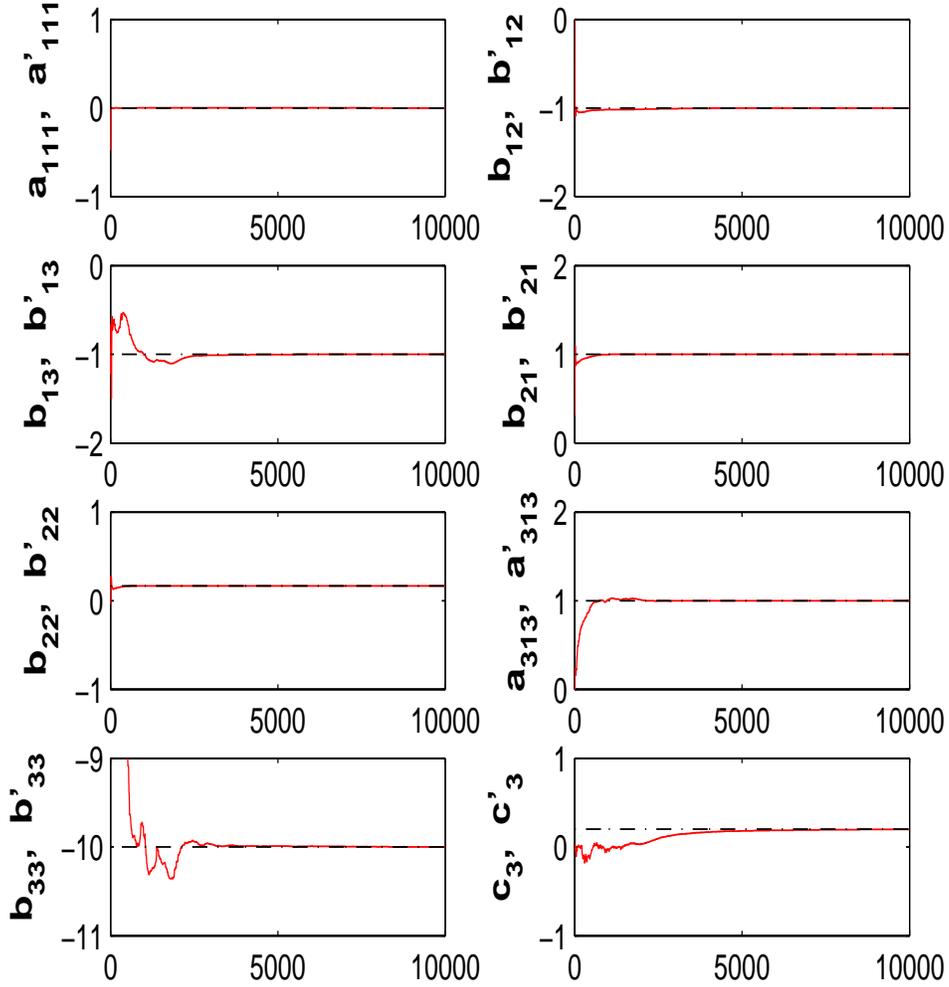,width=14cm,height=15cm}}
\caption{The figure shows the time evolution of some of the estimated parameters for the R\"ossler system (in red), when compared to their true values (black dotted lines), $\nu=10^{-2}/2$, $\gamma=2$.\label{A}}
\end{figure}

\begin{table}[h]
\caption{Coefficients estimated by our strategy for the R\"ossler and the Lorenz systems at the end of a run, $t=10^4$.}
\centering
\begin{tabular}{| l c c | l c c | l c c |}
\hline\hline
 & R\"ossler & Lorenz & & R\"ossler & Lorenz & & R\"ossler & Lorenz
 \\ [0.5ex]
 \hline
  $a'_{111}$ & 0.000 & 0.000 & $a'_{211}$ & 0.000 & 0.000 & $a'_{311}$ & 0.000 & 0.000 \\
  $a'_{112}$ & 0.000 & 0.000 & $a'_{212}$ & 0.000 & 0.000 & $a'_{312}$ & 0.000 & 1.00 \\
  $a'_{113}$ & 0.000 & 0.000 & $a'_{213}$ & 0.000 & -1.00 & $a'_{313}$ & 1.000 & 0.000 \\
  $a'_{122}$ & 0.000 & 0.000 & $a'_{222}$ & 0.000 & 0.000 & $a'_{322}$ & 0.000 & 0.000 \\
  $a'_{123}$ & 0.000 & 0.000 & $a'_{223}$ & 0.000 & 0.000 & $a'_{323}$ & 0.000 & 0.000 \\
  $a'_{133}$ & 0.000 & 0.000 & $a'_{233}$ & 0.000 & 0.000 & $a'_{333}$ & 0.000 & 0.000 \\
  $b'_{11}$ & 0.000 & -10.0 & $b'_{21}$ & 1.000 & 28.0 &   $b'_{31}$ & -0.001 & 0.000 \\
  $b'_{12}$ & -1.00 & 10.0 & $b'_{22}$ & 0.165 & -1.00 &   $b'_{32}$ & -0.001 & 0.000 \\
  $b'_{13}$ & -1.00 & 0.000 &  $b'_{23}$ & 0.000 & 0.000 &  $b'_{33}$ & -10.0 & -2.67 \\
  $c'_{1}$ & 0.000 & 0.000 & $c'_{2}$ & 0.000 & 0.000 &     $c'_{3}$ & 0.202 & 0.000 \\
          \hline
          \end{tabular}
          \end{table}

We have repeated the experiment %presented
above many times, and we have observed either success or failure of our strategy  depending on the random choice of initial conditions for Eqs. (4) and (15). We can explain the failure of our strategy, observed in some experiments,  in terms of
the \emph{transient} dynamics towards synchronization. In fact, if during the identification process, some of the coefficients $a'_{ijk}, b'_{ij}$, and $c'_i$ are not correctly identified,  these may eventually assume values that make ${\bf{x}'}(t)$ too large and ${\bf{x}}(t)$ diverge. Since this divergence typically occurs on a time-scale, which is faster than that on which our strategy operates (and the coefficients are updated), our strategy fails at correctly identify the coefficients. Here, to solve this problem, we propose to replace Eq. (4), by the following equation,
%\begin{equation}
%\dot{\bf{x'}}=\lceil F'({\bf{x'}}) \rceil_{\alpha}+\Gamma(H({\bf{x}})-H({\bf{x'}})), \label{dyn2}
%\end{equation}
\begin{equation}
\dot{\bf{x'}}=\tilde{ F'}({\bf{x'}}) +\Gamma(H({\bf{x}})-H({\bf{x'}})), \label{dyn2}
\end{equation}
where $\tilde{ F'}({\bf{x'}}) =[\tilde{f}_1({\bf{x}}),\tilde{f}_2({\bf{x}}),...,\tilde{f}_n({\bf{x}})]^T$ and
\begin{align}
\tilde{f}_i({\bf{x}})=\left\{ \begin{array} {ccc} {\alpha,} \quad \mbox{if} \quad {{f}_i({\bf{x}})>\alpha}, \\ {f}_i({\bf{x}}),  \quad \mbox{if} \quad    {|{f}_i({\bf{x}})|<\alpha}, \\ -\alpha,  \quad \mbox{if} \quad    {{f}_i({\bf{x}})<-\alpha}, \end{array} \right.
\end{align}
where $\alpha$ is a given constant.
%$\lceil{\bf y} \rceil_{\alpha}= \max{(\min{({\bf y}, \alpha {\bf{1}})}, -\alpha {\bf{1}})}$, and ${\bf{1}}$ is the unitary $n$ dimensional vector column. Note that in so doing, we constrain the components of ${\bf{x'}}$ to lie in the bounded interval $[-\alpha,+\alpha]$.
To test this proposed alternative scheme, we have performed numerical simulations, involving integration of Eq. (17), (13), and (15). For example, by setting a value of $\alpha=10^4$ in Eqs. (17-18), our adaptive strategy was \emph{always} observed to be successful in yielding the correct identification of the parameters.

%We wish to emphasize that our experiments show a certain degree of sensitivity of our strategy with respect to the choice of the initial conditions for Eqs. (4) and (15) and to the transient dynamics toward synchronization. Here we have assumed that that our knowledge of the system to be identified (1) is  limited and therefore we have proceeded under the assumption that the initial conditions for Eqs. (4) and (15) are randomly chosen.  Our hope is that as $t$ increases, all the set of parameters and states are correctly identified and the potentials (5) converge to zero. In this context, the transient dynamics assume a great relevance. In fact the success of our strategy may be compromised by the eventual divergence of some of the states or parameters during the transient towards synchronization.
%A possible countermeasure consists in constraining the state derivatives from becoming too large as $t$ increases, if we consider that no one of the system states/parameters can be infinite. Our experiment also show that an appropriate choice of $\nu$ and $\gamma$ ($\gamma$ should be chosen in the range corresponding to stability of the synchronous solution when two identical systems are coupled as in (4)) can have an impact on the chances for our strategy to work properly.

\subsection{An experiment with the Lorenz system}

We now consider an example in which the unknown system to be identified is the Lorenz system, described by
$\dot{\bf{x}}=F({\bf{x}})$,  ${\bf{x}}=(x_1,x_2,...,x_3)^T$, i.e., $n=3$.
For the Lorenz system, the coefficients $a_{ijk}(t)$, $b_{ij}(t)$, and $c_i(t)$ in Eq. (1) are zero, except for $a_{213}=-1$, $a_{312}=1$, $b_{11}=-10$, $b_{12}=10$, $b_{21}=28$, $b_{22}=-1$, $b_{33}=-8/3$, for which the system is chaotic.

Again, we assume that ${\bf{x}}(0)$  is initialized at a random state on the chaotic attractor, while the system in (\ref{G2}) is evolved starting from a perturbed initial condition (\ref{IC}), with $\rho_1=7.93$, $\rho_2=9.01$, $\rho_3=8.62$. In our test, we use our adaptive strategy described in this paper, with the aim of identifying all the unknown system parameters, $a_{ijk}$, $b_{ij}$, $c_i$. We set $\gamma=10$, and $\nu=10^{-2}$, so that the sliding exponential time $<...>_{\nu}$ is performed over a time $\nu^{-1}=100$ that is long compared to the characteristic time ${T'}_s\simeq 0.97$ for the evolution of a Lorenz system (defined as the time at which the autocorrelation function of $x_1(t)$ decays at one half of its value at $t=0$). Table I shows a comparison of all the $M=30$  parameters estimated at the end of the run ($t=10^4$) to their true values. As can be seen, by using our strategy, we are able to  correctly identify all the $M=30$ unknown parameters of the chaotic Lorenz system.

%\begin{figure}[h]
%\centerline{\psfig{figure=IDTOUT/Riass_Lore2.eps,width=14cm,height=15cm}}
%\caption{The figure shows the time evolution of some of the estimated Lorenz parameters (in red), when compared to their true values (black dotted lines). $\nu=10^{-2}$, $\gamma=10$. \label{B}}
%\end{figure}

\subsection{An experiment with time varying system parameters}

We now present results illustrating the use of
%Since the potential is defined in (\ref{P}) as a \emph{sliding} exponential average of the squared synchronization error, %The reason for introducing a forgetting factor in the formulation of the potential is
our adaptive strategy to dynamically estimate the evolutions of the true system parameters when these are slowly time varying; where by `slowly' we mean that they evolve on a time scale which is much longer than  the averaging time $\nu^{-1}$. %, over which the sliding average (\ref{P}) is defined.% parameters encompass the case where some of the parameters of the true system might be slowly drifting in time.

In this section, we consider that  $\dot{\bf{x}}=F(t,{\bf{x}})$, with ${\bf{x}}=(x_1,x_2,...,x_n)^T$ and $F(t,{\bf{x}})=[f_1({t,\bf{x}}),f_2({t,\bf{x}}),...,f_n({t,\bf{x}})]^T$, where
\begin{equation}
f_i(t,{\bf{x}})=\sum_{j=1}^{n} \sum_{k=j}^{n} a_{ijk}(t) x_j x_k+ \sum_{j=1}^{n} b_{ij}(t) x_j +c_i(t).  \label{TO}
\end{equation}
We expect our identification strategy to work, provided that the averaging time $\nu^{-1} \ll \theta$, where $\theta$ is the time scale on which the true system parameter drift takes place (for a similar use of our adaptive strategy, see \cite{SOTT}). To perform a numerical test (2), we have modified the R\"ossler equations as follows,
\begin{equation}
F({\bf{x}})=\left[\begin{array}{c}
    -x_{2}-x_{3} \\
    x_{1} + 0.165 x_{2} \\
    0.2+ (x_{1}-10+\sin(\omega t) )  x_{3}
  \end{array} \right], \label{FF}
\end{equation}
i.e., all the parameters are the same as in Sec. III.A (i.e., time invariant), but $b_{33}(t)=10+\sin(\omega t)$. System (\ref{FF}) is observed to be chaotic in the range $9 \leq b_{33} \leq 11$. %We take $\omega=(100 T_s)^{-1}$, so as to ensure the parameter drift occur on a time scale which is much slower (one hundred times slower) than the characteristic time scale of an uncoupled system. Therefore,  we can assume that the dynamics of  system (\ref{FF}) remains chaotic at anytime.
%We now assume that the parameter of the true system might be a function of time.

In our test, we use our adaptive strategy described in this paper, Eqs. (4,13,15), with the aim of identifying all the unknown system parameters, $a_{ijk}(t)$, $b_{ij}(t)$, $c_i(t)$. We set $\gamma=2$, and $\nu=0.1$, so that $T_s<\nu^{-1}=10\ll \omega^{-1}$. %, where $\theta$ is assumed to be  the time scale for one oscillation of $b_{33}(t)$, i.e. $\theta=2 \pi \omega^{-1}\simeq 3770$.

%By using our strategy, we are able to  correctly identify all the $M=30$ unknown time-varying parameters.
For the case $\omega= (100 T_s)^{-1}$, Fig. 3 shows the time evolutions of $a'_{313}(t)$ when compared to $a_{313}(t)$, and $b'_{33}(t)$ when compared to $b_{33}(t)$. The estimated parameters are seen to accurately reproduce the evolutions of the true ones. Fig. 4 shows the time averaged identification error ${E}^*_I= \frac{1}{\Delta} \int_{10^4- \Delta}^{10^4} |{b'}_{33}(t)-b_{33}(t)| dt$ ($\Delta=0.3 \times 10^4$) versus  ${\omega} T_s$ for $\gamma=2$, $\nu=0.1$. The point indicated with a full dot corresponds to ${\omega} T_s=0.01$ (the case shown in Fig.3); as can be seen, the identification error becomes large if $\omega T_s \ll 1$ is not satisfied.

\begin{figure}[h]
\centerline{\psfig{figure=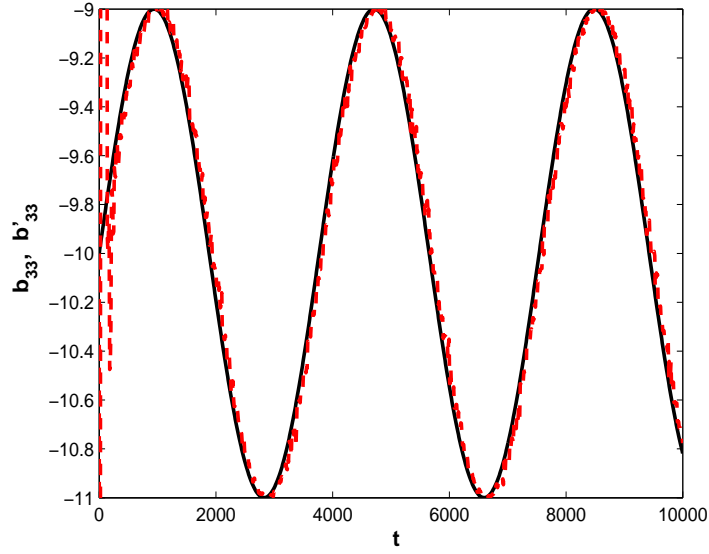,width=10cm,height=8cm}}
\caption{The figure shows the comparison between the time evolutions of two estimated parameters for system (\ref{FF}) (in red), when compared to their true values (black dotted lines), $\omega T_s=0.01$, $\nu=0.1$, $\gamma=2$. In the example, the true  $b_{33}(t)$ is a function of time, i.e., $b_{33}(t)=-10+\sin(\omega t)$. }
\end{figure}

\begin{figure}[h]
\centerline{\psfig{figure=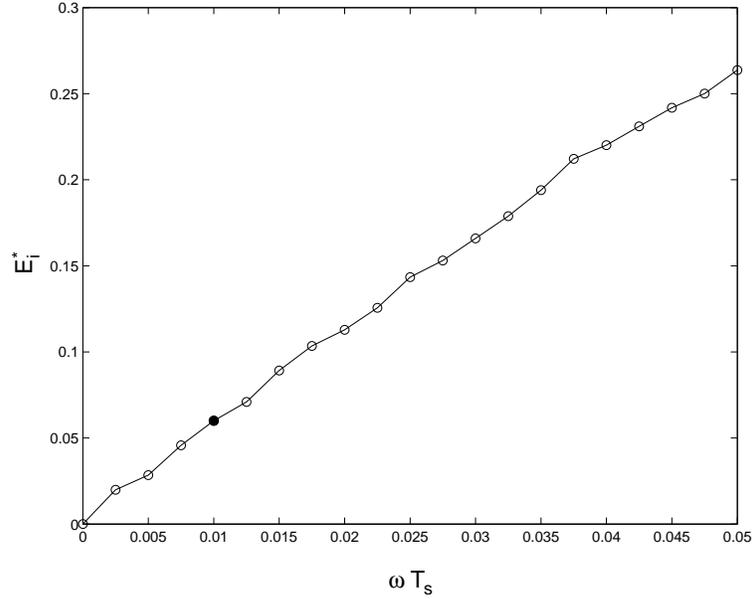,width=10cm,height=8cm}}
\caption{The figure shows the time averaged identification error ${E}^*_I$ versus  $\omega T_s$ for $\gamma=2$, $\nu=0.1$. The point indicated with a full dot corresponds to $\omega T_s =0.01$ (the case shown in Fig. 3). }
\end{figure}

\section{NOISE ANALYSIS}

Here we consider the effects of measurement noise on our identification procedure. Namely we replace Eq. (\ref{dyn}) by
 \begin{equation}
\dot{\bf{x'}}=F'({\bf{x'}})+\Gamma(H({\bf{x^n}})-H({\bf{x'}})), \label{dynn}
\end{equation}
where the noisy state ${\bf{x^n}}=(x^n_1,x^n_2,...,x^n_n)^T$ is equal to ${\bf{x^{n}}}={\bf{x}}+\eta \xi$, $\eta$ is a scalar gain and $\xi=[\xi_1,\xi_2,...,\xi_m]^T$ is an $m$-dimensional vector, whose components are normalized noise terms; we choose  $\xi_i(t)=z \rho_i \epsilon_i(t)$, where for each $i$, $\rho_i$ is the standard deviation of the time evolution of $x_i$ for the true system; $\epsilon_i(t)$ at each time step of our numerical integration are zero-mean independent random numbers of unit variance drawn from a normal distribution; $z$ is normalization factor, which
is chosen so that $\eta \sim 1$ makes the noise cause the relevant state component to
diffuse by an amount that, over a time interval $T_s$, is roughly as big as the amplitude of
variation of the relevant chaotic state component, and is given by $z=\sqrt {\frac{\tau}{T_s}}$, where $\tau$ is the time step of our integration method and $T_s$ is the period of one oscillation of the true system.

Measurement noise introduces a rapid small-scale variation to the measurement time series $\bf{x}^n(t)$, thus making the meaning of the time derivative $d {\bf{x}^n(t)}/dt$ that would appear in  Eq. (\ref{P}) questionable. Therefore we replace Eq. (\ref{P}) by
\begin{equation}
\Psi_i=<[\dot{x^{*}}_i-f'_i(x'_1,x'_2,...,x'_n)]^2>_{\nu}, \qquad i=1,2,...,n,   \label{P2}
\end{equation}
where we obtain $x^*_i(t)$  by low pass filtering the noisy state $x^n_i(t)$, with cutoff frequency $T$. %, that is $x^*_i(t)=\frac{1}{T}\int^t e^{(t-u)/{T}} x_i^n(u) du$.
Then  $\dot{x^{*}}_i$ to be inserted in (\ref{P2}) is obtained from the differential equation
\begin{equation}
\dot{x^{*}}_i(t)+\frac{1}{T} x^*_i(t)=\frac{1}{T} x^n_i(t),
\end{equation}
where  $T$ is chosen to satisfy $T_s >> T >> \tau$. In our numerical experiment, we choose $\tau=10^{-3}$, and $T= 4 \times 10^{-3}$.

We monitor the robustness of our identification strategy with respect to increasing values of the noise $\eta$. To this aim, we introduce the following identification error measure, %two error measures:
%\begin{equation}
%E_{S}(t)=\frac{1}{n} \sum \frac{1}{\rho_i} { |x_{i}(t)-{x'}_{i}(t)|} ,
%\end{equation} and,
\begin{equation}
E_{I}(t)=\frac{1}{M} [ { \sum_{i=1}^{n} \sum_{j=1}^{n} \sum_{k=j}^n { |{a}'_{ijk}(t)-a_{ijk}|} + \sum_{i=1}^{n} \sum_{j=1}^{n} { |{b}'_{ij}(t)-b_{ij}|} + \sum_{i=1}^{n} { |{c}'_{i}(t)-c_i|} } ]. \label{IE}
\end{equation}

We have performed numerical tests to investigate how %the time averaged synchronization errors $E_S= \frac{1}{\Delta} \int_{ 10^{4}- \Delta}^{10^4} E_S(t) dt$ and
the time averaged identification error $\bar{E}_I= \frac{1}{\Delta} \int_{10^4- \Delta}^{10^4} E_I(t) dt$ depends on $\eta$ ($\Delta=0.3 \times 10^4$). Our simulations show that $\bar{E}_I$ remains small through the range tested, and for example is less than $1 \%$ in the range $0 \leq \eta \leq 1$, when $\nu=4 \times 10^{-2}$, $\gamma=5$.
%\begin{figure}[h]
%\centerline{\psfig{figure=IDTOUT/Meas_Noise.eps,width=10cm,height=6cm}}
%\caption{The figure shows the time averaged identification error $E_I$ versus increasing noise $\eta$, $\nu=4 \times 10^{-2}$, $\gamma=5$.\label{Meas_Noise}}
%\end{figure}

\section{THE CASE OF INCONSISTENT MODEL EQUATIONS}

In this section, we suppose that the dynamics of the true system deviates from polynomial form (i.e., $f_i(\bf{x})$  cannot be written as a given degree polynomial in $\bf{x}$), but we still do our adaptive procedure using a polynomial model as in Eq. (\ref{TO}). As a first attempt, we continue to assume that the system dynamics can be approximately modeled as a degree 2-polynomial as in (\ref{G2}) and in so doing we want to test the limits of this approach. Let us consider for example that the true system dynamics is described by $\dot{\bf{x}}=F({\bf{x}})$, where $F({\bf{x}})$ is given by,
\begin{equation}
F({\bf{x}})=\left[\begin{array}{c}
    -x_{2}-x_{3} + \kappa \rho_1 sech(x_1) \\
    x_{1} + 0.165 x_{2} \\
    0.2+ (x_{1}-10 )  x_{3}
  \end{array} \right]. \label{RC}
\end{equation}
Note that for $\kappa=0$, Eq. (\ref{RC}) is equivalent to the R\"ossler Eq. (\ref{F}), while for $\kappa \neq 0$ the system dynamics includes a transcendental function, i. e., a function that cannot be expressed as a finite degree polynomial. In what follows we implement the same adaptive strategy described in Sec. II and test its effectiveness for increasing values of $\kappa$.

In following our adaptive procedure, because the true and model systems are different, the evolving model parameters from our system of adaption equations do not settle into constant values. Rather we observe that they time-asymptotically fluctuate around some constant value. Thus we take for our identification of the model parameters the time average of these quantities.

In order to formulate a measure of the effectiveness of our identification procedure with an inconsistent model, we note that for practical purposes, one is often  interested in  how well a model is able to reproduce the behavior of the real system. In
particular, a sensible question to ask would be how well the model equations obtained through our adaptive strategy forecast the future behavior of the true system, and, in particular how far in the future are such forecasts reliable. In what follows we provide a partial answer to such a question.

\begin{figure}[h]
\centerline{\psfig{figure=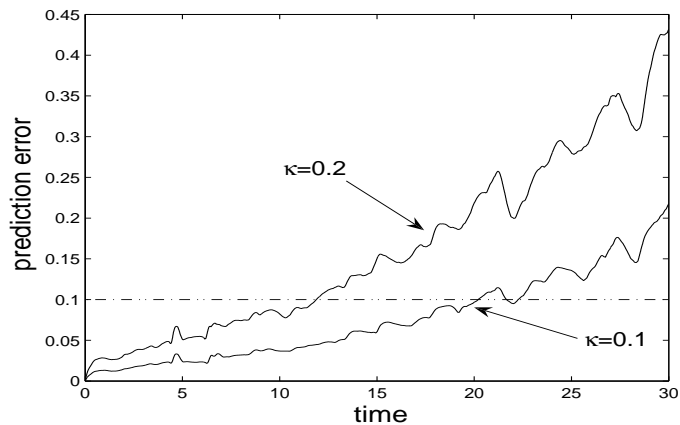,width=10cm,height=6cm}}
\caption{The figure shows the prediction error ${\rho_1}^{-1} |x_1(t)-x'_1(t)|$ as function of time for the cases of $\kappa=0.1$  and $\kappa=0.2$. The results have been averaged over $500$ different choices of the initial conditions. \label{pred}}
\end{figure}

We have performed numerical experiments in which we evaluate the error of the obtained model system when it is used to forecast the evolution of the true system. We consider two cases, $\kappa=0.1$ and $\kappa=0.2$. %All the estimated coefficients less than $10^{-2}$ in modulus have been set to $0$.
In Fig. \ref{pred}  we have monitored the forecast error ${\rho_1}^{-1} |x_1(t)-x'_1(t)|$ as function of $t$, when the model and the system are uncoupled and evolved from the same initial condition $x_1(0)=x'_1(0)$ (which we have taken to be a randomly chosen point from the R\"ossler attractor). The results in Fig. \ref{pred}  have been averaged over $500$ different choices of the initial conditions. From Fig. \ref{pred}, we observe that for example, if we set our prediction time to be  the length of time over which the prediction error remains less than 10\%, it is about $12$ (about $20$) in the case of  $\kappa=0.1$ ($\kappa=0.2$). In particular, the latter time is more than three times the characteristic time scale of the true system $T_s \simeq 6$.

%
%\begin{figure}[h]
%\centerline{\psfig{figure=IDTOUT/EiEs.eps,width=10cm,height=6cm}}
%\caption{The figure shows the time averaged identification error $E_I$ (solid line) and synchronization error $E_S$ (dashed line) versus increasing values of the parameter $\kappa$, $\nu=0.01$, $\gamma=5$.\label{EIES}}
%\end{figure}

%\begin{table}[h]
%\caption{Average value and standard deviation of the estimated coefficients at the end of a run.}
%\centering
%\begin{tabular}{| l | l l |  l l |  l l |}
%\hline\hline
% & {mean}($\kappa$=0) & {std} ($\kappa$=0)  & mean ($\kappa$=0.5) & std ($\kappa$=0.5)& mean ($\kappa$=1)& std ($\kappa$=1)
% \\ [0.5ex]
% \hline
%  $a'_{111}$ & 0.000 & 0.000 & -0.0092 & 0.000145 & -0.0165 &   0.000163  \\
%  $a'_{112}$ & 0.000 & 0.000 & -0.0003 & 0.000101 &  0.00853 &  0.000312  \\
%  $a'_{113}$ & 0.000 & 0.000 & -0.0063 & 0.000345 &  -0.0692 &  0.00127  \\
%  $a'_{122}$ & 0.000 & 0.000 &  0.0001&  0.000160 &  -0.00113 & 0.000173 \\
%  $a'_{123}$ & 0.000 & 0.000 & -0.0435 & 0.00105 &  -0.302 &   0.00496  \\
%  $a'_{133}$ & 0.000 & 0.000 & -0.0012 & 0.000727 &  0.00815 &  0.000184  \\
%  $b'_{11}$ & 0.000 & 0.000 &   0.0043 & 0.00107 &  0.0652 &    0.00297  \\
%  $b'_{12}$ & -1.00 & 0.000 &  -0.976 & 0.000946 &  -0.893 &   0.00221  \\
%  $b'_{13}$ & -1.00 & 0.000 &  -0.867 & 0.00458 &  -0.888 &  0.00510  \\
%  $c'_{1}$ & 0.000 & 0.000 &    1.19 &  0.0284 & 3.24 &     0.0376  \\
%          \hline
%          \end{tabular}
%          \end{table}
%%tempo 10000, nu=10^-2/2,

\section{CONCLUSION}

In this paper, we have introduced a new strategy to identify the  parameters of an unknown chaotic dynamical system. We aim at synchronizing the real unknown system with another system \emph{`in silico'}, whose parameters are adaptively evolved to converge on those of the real one.

As a first attempt, we have assumed that the differential equations governing the system dynamics are expressible, or approximately expressible, in terms of polynomials of an assigned degree. For this case, our strategy relies on the assumption that the only necessary information about the true system is the dimensionality of its state vector and the order of the polynomials.
Under these conditions, we have shown that we are able to extract the whole set of parameters of the unknown system from knowledge of the dynamical evolution of its state vector and its first derivative. Our procedure relies on the minimization of appropriately defined potentials that are zero when both the system state and parameters are correctly identified. Interestingly, our strategy is effective in detecting which parameters are/are not zero%(and thus do not affect the system dynamics)
, and in obtaining correct estimates for those that are not zero.

We have further considered the effects of measurement noise and we have proposed an alternative scheme that works when only knowledge of the dynamical evolution of true system state vector is available, which has been shown to be effective even in the presence of relatively high noise. %Our results show that our procedure is quite robust with respect to measurement noise.

We have also analyzed a situation in which the model fitting function basis is slightly inconsistent with the true system dynamics, and, for this case, we have evaluated how well the obtained model is able to forecast the future behavior of the true system, i.e., how far into the future it is able to forecast its evolution.

As a further application, we presented the possibility of extending our approach to the case in which the parameters to be identified are slowly varying in time (i.e., on a time scale that is slower than $\nu^{-1}$). The general strategy can also be used if one has a system of known form with several unknown parameters.

This work was supported by the U.S. Office of Naval Research, contract N00014-07-1-0734.

\end{document}